\documentstyle[prl,floats,epsfig,aps]{revtex}
\input{psfig.sty}
\begin{document}

\title{Exponentially hard problems are sometimes polynomial,\\ 
a large deviation analysis of search
algorithms for the random Satisfiability problem, \\
and its application to stop-and-restart resolutions.}

\vskip .5cm
\author{Simona Cocco $^{1}$ and R{\'e}mi Monasson $^{2}$}
\address{$^1$  CNRS-Laboratoire de Dynamique des Fluides Complexes,
3 rue de l'Universit{\'e}, 67000 Strasbourg, France.\\
$^{2}$ CNRS-Laboratoire de Physique Th{\'e}orique de l'ENS,
24 rue Lhomond, 75005 Paris, France.}
\date{\today}
\maketitle  

\begin{abstract}
A large deviation analysis of the solving complexity of random
3-Satisfiability instances slightly below threshold is
presented. While finding a solution for such instances demands an
exponential effort with high probability, we show that an
exponentially small fraction of resolutions require a computation
scaling linearly in the size of the instance only. This exponentially
small probability of easy resolutions is analytically calculated, and
the corresponding exponent shown to be smaller (in absolute value)
than the growth exponent of the typical resolution time. Our study
therefore gives some theoretical basis to heuristic stop-and-restart
solving procedures, and suggests a natural cut-off (the size of the
instance) for the restart.
\end{abstract}
\vskip .5cm
Computational problems are usually divided into two classes. They are
easy if there exists a solving procedure whose running time 
grows at most polynomially with
the size of the problem, or hard if no such algorithm is believed
to exist, and the best avalaible procedures may require an exponentially
growing time\cite{papa}. The polynomial vs. exponential classification 
was enriched in the last decades through the
derivation of quantitative bounds on resolution complexity, and 
the study of average performances of various resolution algorithms 
for computational tasks with model input distributions.

In this letter, we show that, though the polynomial/exponential 
dichotomy certainly applies to the typical resolution complexity
of computational problem, it may not be so for 
large deviations from typical behavior. Typically exponentially hard
problems may sometimes be solved in polynomial time, a phenomenon we
take advantage of to accelerate resolution drastically.

We concentrate here on random 3-Satisfiability (3-SAT), a paradigm of
hard combinatorial problems recently studied using statistical physics
tools and concepts \cite{Sta,Coc} as e.g. number
partitioning\cite{Mertens}, vertex cover\cite{Weigt}, ... to which
computer scientists have devoted a great attention over the past
years\cite{survey,Mit,runtorun}.  An instance of random 3-SAT is
defined by a set of $M$ constraints (clauses) on $N$ boolean ({\em
i.e.} True or False) variables. Each clause is the logical OR of three
randomly chosen variables, or of their negations.  The question is to
decide whether there exists a logical assignment of the variables
satisfying all the clauses (called solution). The best currently known
algorithm to solve 3-SAT is the Davis--Putnam--Loveland--Logemann
(DPLL) procedure\cite{survey} (Fig.~1). The sequence of assignments of
variables made by DPLL in the course of instance solving can be
represented as a search tree (Fig~2), whose size $Q$ (number of nodes)
is a convenient measure of the complexity.  For very large sizes ($M,N
\to \infty$ at fixed ratio $\alpha=M/N$), some static and dynamical
phase transitions arise\cite{Sta,Mit,runtorun,Kir,Friedgut}.  Instances
with a ratio of clauses per variable $\alpha >\alpha _C \simeq 4.3$
are almost surely 
(a.s.) unsatisfiable and obtaining proofs of refutation require an
exponential effort\cite{Coc,Chv}. Below the static threshold $\alpha
_C$, instances are a.s. satisfiable, but finding a solution may be
easy or hard, depending on the value of $\alpha$. A dynamical
transition \cite{Coc,Vardi} takes place at $\alpha_L\simeq 3.003$ (for
the heuristic used by DPLL shown in Fig.~1) separating a polynomial
regime ($\alpha < \alpha_L : \; Q\sim N$, search tree A on Fig.~2)
\cite{Fra,Fri} from an exponential regime ($\alpha >\alpha _L: Q \sim
2^{N\omega}$, search tree B).  This pattern of complexity, and the
value of $\omega (\alpha)$ were obtained through an analysis of DPLL
dynamics, reminiscent of real-space renormalization in statistical
physics\cite{Coc}.  DPLL generates some dynamical flow of the
instance, whose trajectory lies in the phase diagram of the 2+p-SAT
model\cite{Sta}, an extension of 3-SAT, where $p\le 1$ is the fraction
of 3-clauses (Fig.~2).

We focus hereafter on the large deviations of complexity 
in the upper sat phase $\alpha _L < \alpha < \alpha _C$. Using
numerical experiments and analytical calculations, we show that,
though complexity $Q$ a.s. grows as $2^{N\omega}$,
there is a finite, but exponentially small, probability $2^{-N
\zeta}$ that $Q$ is bounded from above by $N$ only.  In other words,
while finding solutions to these sat instances is almost always
exponentially hard, it is very rarely easy (polynomial time). Taking
advantage of the fact that $\zeta$ is smaller than $\omega$, we show
how systematic restarts of the heuristic may decrease substantially
the overall search cost. Our study therefore gives some theoretical
basis to stop--and--restart solving procedures empirically
known to be efficient\cite{Dubois}, and suggests a natural
cut-off for the stop.

% numerical experiments

Distributions of resolution times $Q$ for $\alpha =3.5$ are reported 
on Fig.~3. The histogram of $\omega=(\log_2 Q)/N$ essentially
exhibits a narrow peak (left side) followed by a
wider bump (right side). As $N$ grows, the right peak acquires more
and more weight, while the left peak progressively disappears.
The center of the right peak gets slightly
shifted to the left, but  reaches a finite value 
$\omega ^* \simeq 0.035$ as $N\to \infty$\cite{Coc}. This right peaks
thus corresponds to the core of exponentially hard resolutions: 
resolutions of instances a.s. require a time scaling as 
$2^{N \omega ^*}$ as the size $N$ gets large,
in agreement with the above discussion.

On the contrary, the abscissa of the maximum of the left peak vanishes
as $\log_2 N/N$ when the size $N$ increases, indicating that the left
peak accounts for polynomial (linear) resolutions. Its maximum is
located at $Q/N\simeq 0.2-0.25$, with weak dependence upon $N$. The
cumulative probability $P_{lin}$ to have a complexity $Q$ less than,
or equal to $N$, decreases exponentially: $P_{lin}= 2 ^{-N \zeta}$
with $\zeta \simeq 0.011 \pm 0.001$ (Inset of Fig.~4).  In the
following we will concentrate on linear resolutions only
(an analysis of the distribution of exponential resolutions for the
problem of the vertex covering of random graphs\cite{Weigt} can be
found in \cite{Mont}).

Further numerical investigations show that, in easy resolutions, the
solution is essentially found at the end of the first branch, with a
search tree of type $A$, and not $B$, in Fig.~2. 
Easy resolution trajectories are able to cross the 'dangerous' region
extending beyond point $D$ in Fig.~2,
contrary to most trajectories which backtrack earlier.
Beyond $D$, unit-clauses (UC) indeed accumulate. Their number $C_1$ 
becomes of the order of $N$ ($C_1/N \simeq 0.022$ for $\alpha
=3.5$), and the probability that the branch survives, {\em i.e.} 
that no two contradictory UC are present, is exponentially small in $N$, in
agreement with the scaling of the left peak weight in Fig.~3.

Calculation of $\zeta$ requires the analysis of the first descent in 
the search tree (Fig.~2). Each time DPLL assigns a variable some clauses 
are eliminated, other are reduced or left unchanged (Fig.~1). We thus 
characterize an instance by its state ${\bf C}=(C_1,C_2,C_3)$, where 
$C_j$ is the number of $j$-clauses it includes ($j=1,2,3$). Initially, 
${\bf C}=(0,0,\alpha_0 N)$. Let us call $\tilde P({\bf C}; T)$ the
probability that the assignment of $T$ variables has produced
no contradiction and an instance with state ${\bf C}$.  $\tilde P$
obeys a Markovian evolution $\tilde P({\bf C}; T+1) = \sum _{\bf C'} K
({\bf C},{\bf C'}; T) \tilde P({\bf C'}; T)$ where the entries of the
transition matrix $K$ read
\begin{eqnarray}
\label{bbra}
K ({\bf C},{\bf C'}; T) &=& B_{p_3}^{ C_3', \Delta _3}
\sum_{w_3=0}^{\Delta _3} B_{1/2}^{\Delta _3 , w_3} 
\sum_{z_2=0}^{C_2'-v} B_{p_2}^{C_2'-v , z_2 } 
\sum_{w_2=0}^{z_2} B_{1/2}^ { z_2, w_2 }
\nonumber \\&&\times   
\sum_{z_1=0}^{C_1'-1+v}\frac {1} {2^{z_1}}  B_{p_1}^{C_1'-1+v,z_1 } 
\; \delta_{z_2-\Delta _2-w_3+v}\;\delta_{z_1-\Delta _1-w_2+1-v}  
\end{eqnarray}
where $\delta _C$ denotes the Kronecker function: $\delta _C=1$
if $C=0$, $0$ otherwise. Variables appearing in (\ref{bbra}) are as
follows. $\Delta _j\equiv C_j'-C_j$, $v\equiv\delta_{C'_1}$, 
$z_j$ (respectively $w_j$) is the number of $j$-clauses
which are satisfied (resp. reduced to $j-1$ clauses) when the
$(T+1)^{th}$ variable is assigned. These are stochastic variables drawn
from several binomial distributions $B_p ^{L , K}\equiv{L \choose
K}p^{K} (1-p)^{L-K}$.  Parameter $p_j= j/(N-T)$ equals the probability
that a $j$-clause contains the variable just assigned by DPLL.

The introduction of the generating function $P({\bf y};T) 
=\sum_{\bf C}\; e^{\,{\bf y} \cdot {\bf C}}\;\tilde P({\bf C},T)$, 
allows us to express the evolution equation for the state 
probabilities in a compact manner,
\begin{equation}
\label{eqev}
{P}(\,{\bf y}\,;T+1\,)=e^{-g_1({\bf y})}\;{P}\big(
\,{\bf g}({\bf y})\,; T\,\big)+
\left ( e^{- g_2({\bf y})}- e^{- g_1({\bf y})}\right)  
{P}\big(-\infty,\,g_2({\bf y}),\,g_3({\bf y})\,; T\, \big)
\end{equation}
where $g_j({\bf y})=y_j+\ln (1+\gamma_j({\bf y}) /{N})$,
$\gamma_j ({\bf y})\equiv\gamma_j (y_j,y_{j-1})=j \,(e^{-y_j}
(1 +e^{y_{i-1}})/2 -1)/(1-t)$ for $j=1,2,3$  ($y_0 \equiv
-\infty$).

From (\ref{bbra}),the $C_j$s undergo $O(1)$ changes each time a
variable is fixed. After $T = t\, N$ assignments, the densities
$c_j=C_j/N$ of clauses have been modified by $O(1)$. This translates
into large $N$  Ans{\"a}tze for the state probability, $\tilde
P({\bf C};T) =e^{N \varphi({\bf c};t)}$, and for the generating
function, $P({\bf y};T ) = e ^{\, N \, \varphi ( \,{\bf y}\, ; t\,)}$,
up to non exponential in $N$ terms. $\varphi$ and $\tilde \varphi$ are
simply related to each other through a Legendre transform.  In
particular, $\varphi ({\bf 0};t)$ is the logarithm of the
probability (divided by $N$) that the first branch has not been hit by
any contradiction after a fraction $t$ of variables have been
assigned. The most probable values of the densities $c_j(t)$ of
$j$-clauses are equal to the partial derivatives of
$\varphi$ in ${\bf y}={\bf 0}$.

When DPLL starts running on a 3-SAT instance, clauses are reduced and
some UC generated. Next they are eliminated through 
UC propagation, and splits occur frequently (Fig.~1). The number $C_1$ of
UC remains bounded with respect to the instance size $N$, and
the density $c_1(t)=\partial \varphi/\partial y_1$ identically vanishes.
$\varphi$ does not depend on $y_1$, and $\varphi(y_2,y_3;t)$ obeys 
the following partial differential equation (PDE) 
\begin{equation} \label{pde1}
\frac{\partial \varphi}{\partial t}= -y_2 + \gamma _2 
(y_2,y_2;t)\; \frac{\partial \varphi}{\partial y_2} 
+ \gamma _3 (y_2,y_3;t)\;
\frac{\partial \varphi}{\partial y_3} 
\end{equation}
We have solved analytically PDE (\ref{pde1}) with initial condition 
$\varphi({\bf y};0) = \alpha _0 y_3$. The high probability scenario
is obtained for $y_2=y_3=0$: $\varphi (0,0;t)=0$ indicates that 
the probability of survival of the branch is not exponentially small in 
$N$\cite{Fri}, and the partial derivatives $c_2(t), c_3(t)$ give
the typical densities of 2- and 3-clauses, in full agreement with Chao 
and Franco's result\cite{Fra}. We plot in Fig.~2 the corresponding
resolution trajectories for various initial ratios $\alpha _0$,
using the change of variables $p=c_3/(c_2+c_3)$,
$\alpha=(c_2+c_3)/(1-t)$.  
Our calculation provides furthermore a complete
description of rare deviations of the resolution trajectory from its
highly probable locus, giving acces to the  exponentially
small probabilities that $p, \alpha$ differ from their most
probable values at `time' $t$.  

The assumption $C_1= O(1)$ breaks down for the most probable
trajectory at some fraction $t_D$ e.g. $t_D\simeq 0.308$ for $\alpha
_0=3.5$ at which the trajectory hits point $D$ on Fig.~2.
Beyond $D$, UC accumulate, and the probability of survival of
the first branch becomes exponentially small in $N$.
Variables are almost always assigned through unit-propagation:
$c_1 > 0$. $\varphi$ now depends on $y_1$ and, from (\ref{bbra}), obeys 
the following  PDE 
\begin{equation} \label{pde2}
\frac{\partial \varphi}{\partial t} = -y_1 + \sum _{j=1} ^3\gamma _j 
({\bf y};t)\; \frac{\partial \varphi}{\partial y_j} 
\end{equation}
We have solved PDE (\ref{pde2}) through an expansion of $\varphi$ in
powers of ${\bf y}$, whose coefficients obey, from (\ref{pde2}), a set 
of coupled linear ODEs. The initial conditions for the
ODEs are chosen to match the expansion of the exact solution of
(\ref{pde1}), that is, the typical trajectory and its large
deviations, at time $t_D$. The quality of the approximation improves 
rapidly with the order $k$ of the expansion, and no difference was 
found between $k=3$ and $k=4$ results.  $c_1$ first increases, 
reaches its top value $(c_1) ^{max}$, then decreases
and vanishes at $t_{D'}$ when the trajectory comes out
from the "dangerous" region  where contradictions a.s. occurs
(Fig.~2). The probability of survival scales as $2^{- N \zeta}$ 
for large $N$, with $\zeta  = -\varphi( {\bf 0};t_{D'})/\ln 2 $.  
The calculated values of $\zeta 
\simeq 0.01, (c_1)^{max}\simeq 0.022$ and $Q/N \simeq 0.21$ for
$\alpha=3.5$ are in very good agreement with numerics. Fig.~4 shows
the agreement between theory and simulations over the
whole range $\alpha _L < \alpha < \alpha _C$.

The existence of rare but easy resolutions suggests the use of a
systematic stop-and-restart (S\&R) procedure to speed up resolution: if a
solution is not found before $N$ splits, DPLL is stopped and rerun
after some random permutations of the variables and clauses.  The
expected number $N_{rest}$ of restarts necessary to find a solution
being equal to the inverse probability $1/P_{lin}$ of linear resolutions,
the resulting complexity should scale as $N\, 2^{\,0.011\,N}$ for
$\alpha=3.5$, with an exponential gain with respect to DPLL one-run 
complexity, $2^{\, 0.035 \, N}$. Results of S\&R experiments are
reported on Fig~4. The typical 
number $N_{rest}=2^{N\bar \zeta}$ of restarts  
grows indeed exponentially with the size $N$, with a rate
$\bar \zeta = 0.0115 \pm 0.001$ equal to $\zeta$ \cite{nota}.   
Performances are greatly enhanced by the use of S\&R (see Fig.~4
for comparison between $\zeta$ and $\omega$).
While with usual DPLL, we were able to solve instances with 500
variables in about one day of CPU for $\alpha =3.5$,  
instances with 1000 variables were solved with S\&R in 15 minutes 
on the same computer.

Our work therefore provides some theoretical support to the use of 
S\&R\cite{Dubois,Mont}, and in addition suggests a natural cut-off at
which the search is halted and restarted, the determination of which is 
usually widely empirical and problem dependent. If a combinatorial
problem is efficiently solved (polynomial time) by a search 
heuristic for some values of the control parameter of the input distribution,
there might be an exponentially small probability that the heuristic
is still successfull (in polynomial time) in the range of parameters
where resolution almost surely requires massive backtracking and exponential
effort. When the decay rate of the polynomial time 
resolution probability $\zeta$ is smaller than the growth rate
$\omega$ of the typical exponential resolution time, S\&R 
with a cut-off in the search equal to a polynomial of the 
instance size will lead to an exponential speed up of resolutions.

{\bf Acknowledgments.} 
We thank A. Montanari and R. Zecchina for discussions and communication
of their results prior to publication, and the French Ministry of
Research for a partial financial support through the ACI Jeunes Chercheurs 
``Algorithmes d'optimisation et syst{\`e}mes d{\'e}sordonn{\'e}s quantiques''.

\begin{figure}[b]
\begin{center}
\includegraphics[width=340pt,angle=-0]{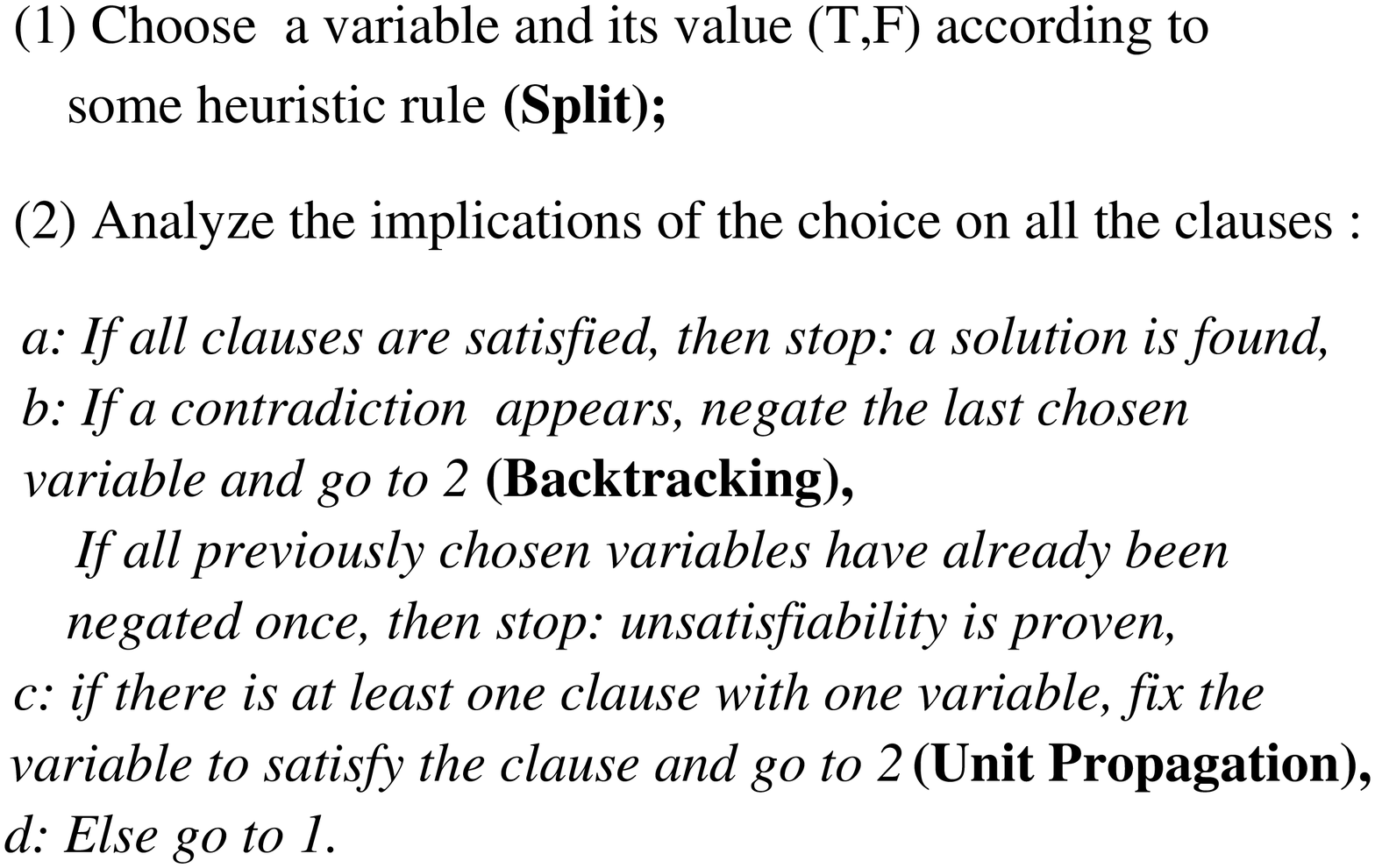}
\end{center}
\vskip .5cm
\caption{DPLL algorithm.  When a variable has been chosen at step (1)
e.g. $x=T$, at step (2) some clauses are satisfied e.g. $C=(x \;\hbox
{\rm OR}\;{y} \;\hbox {\rm OR}\; z)$ and eliminated, other are reduced
e.g. $C=(\hbox{\rm not}\, x\; \hbox {\rm OR}\;{y}\; \hbox {\rm OR}\;
{z}) \to C=({y}\; \hbox {\rm OR}\; {z})$. If some clauses include one
variable only e.g. $C=y$, the corresponding variable is automatically
fixed to satisfy the clause ($y=T$).  This unit--clause (UC)
propagation (2c) is repeated up to the exhaustion of all
UC. Contradictions result from the presence of two opposite UC
e.g. $C=(y), C'=(\hbox{\rm not}\, y)$. A solution is found when no clauses
are left.  The heuristic studied here is the Generalized UC (GUC)
rule: a variable is chosen at step (1) from one of the 2-clauses (or
from a 3-clause if no 2-clause is present), and fixed to satisfy the
clause. The search process of DPLL is represented by a tree (Fig.~2) whose
nodes correspond to (1), and edges to (2).  Branch extremities are
marked with contradictions C (2b), or by a solution S (2a).}
\end{figure}

\begin{figure}
\begin{center}
\includegraphics[width=320pt,angle=-90]{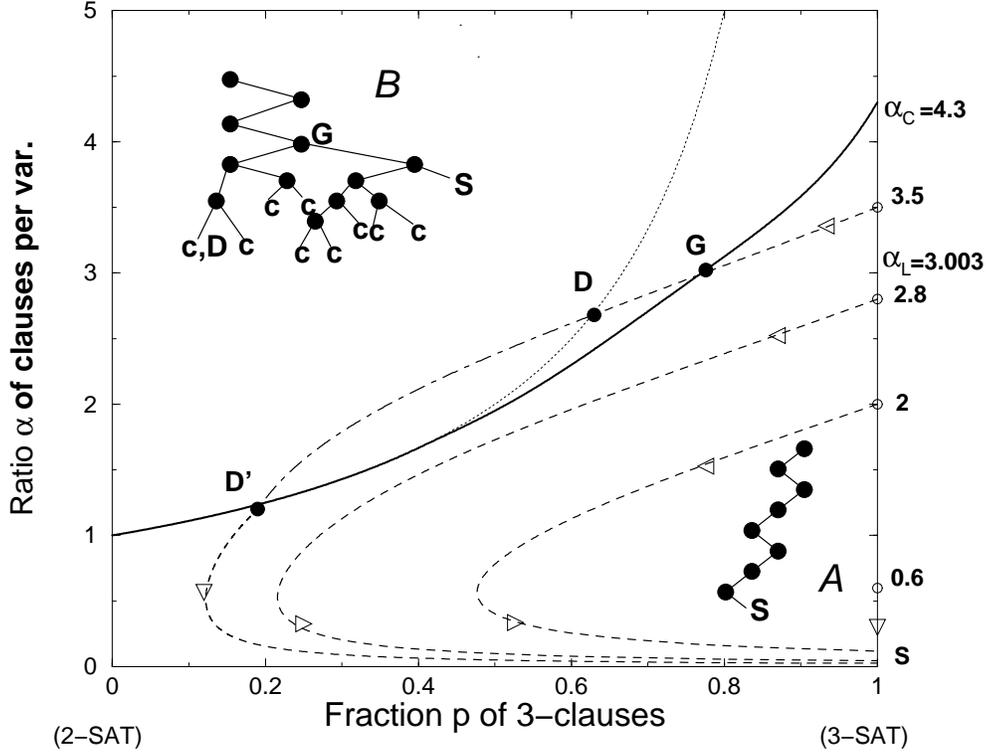}
\end{center}
\vskip .5cm
\caption{Phase diagram of 2+p-SAT and first branch trajectories for 
satisfiable instances. The threshold line $\alpha_C (p)$
(bold full line) separates sat (lower part of the plane) from unsat
(upper part) phases.  Departure points for DPLL trajectories are
located on the 3-SAT vertical axis (empty circles).  Arrows
indicate the direction of "motion" along trajectories (dashed curves)
parameterized by the fraction $t$ of variables set by DPLL.  For small
ratios $\alpha < \alpha _L$ ($\simeq 3.003$ for the GUC heuristic),
branch trajectories remain confined in the sat phase, end in $S$ of
coordinates $(1,0)$, where a solution is found (with a search process
reported on tree A).  For ratios $\alpha _L < \alpha < \alpha_C$, the
branch trajectory intersects the threshold line at some point
$G$. A contradiction a.s. arises before the trajectory crosses the dotted curve
$\alpha =1/(1-p)$ (point D), and extensive backtracking up to $G$ permits to
find a solution (Search tree B).
With exponentially small probability, the trajectory (dot--dashed
curve, full arrow) is able to cross the "dangerous" region where
contradictions are likely to occur (Search tree similar to A); 
it then exits from this region
(point $D'$) and ends up with a solution (lowest dashed trajectory).}
\label{diag}
\end{figure}

\begin{figure}
\begin{center}
\includegraphics[height=300pt,angle=-90] {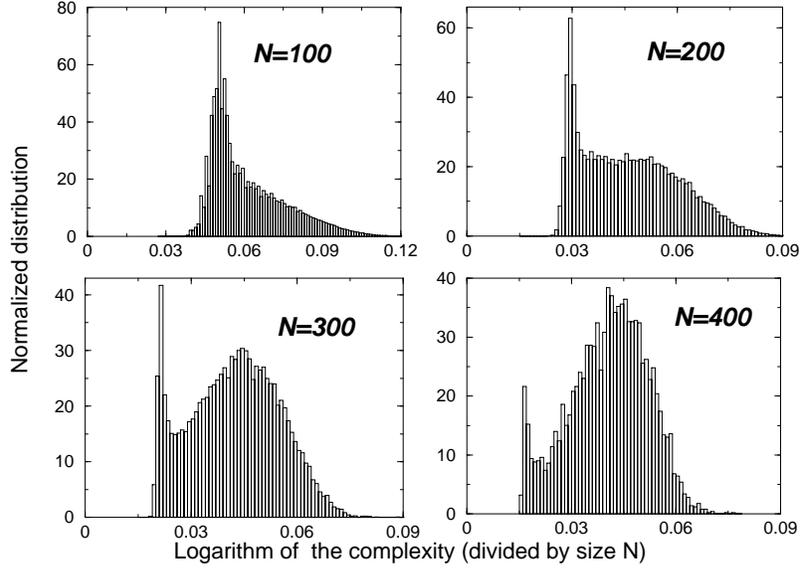}
\end{center}
\vskip .5cm
\caption{Histograms of the logarithm $\omega$ of the complexity $Q$
(base 2, and divided by $N$) for $\alpha=3.5$ and different sizes $N$.
Many instances are drawn randomly, and for each sample, 
DPLL is run until a solution is found 
(very few unsatisfiable instances can be present and are discarded).}
\label{histolog}
\end{figure}

\begin{figure}
\begin{center}
\includegraphics[height=250pt,angle=-90] {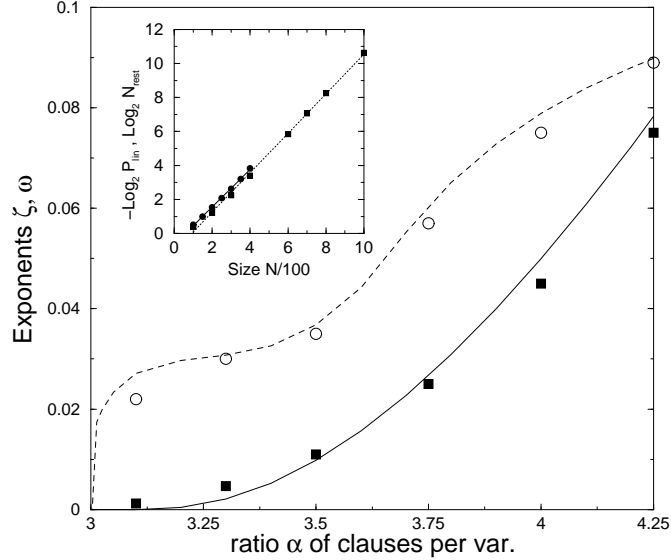}
\end{center}
\caption{Log. of complexity using DPLL ($\omega$ -- simulations:
circles, theory from [3]: dotted line) and S\&R
($\zeta$ -- simulations: squares, theory: full line) 
as a function of ratio $\alpha$. Inset: Minus log. of 
the cumulative probability $P_{lin}$  of complexities 
$Q\le N$ as a function of the size for $100 \le N\le 400$ 
(full line); log. of the number of restarts $N_{rest}$ 
necessary to find a solution for $100\le N \le 1000$ 
(dotted line) for $\alpha=3.5$. Slopes are $\zeta = 0.0011$ and $\bar
\zeta = 0.00115$ respectively.}
\label{histolin}
\end{figure}

\end{document}